\documentclass[11pt,letterpaper,notoc]{JHEP}   
\usepackage{epsfig}
\newcommand{\half}{\frac12}

\newcommand{\beq}{\begin{eqnarray}}
\newcommand{\eeq}{\end{eqnarray}}

\def\CO{{\cal O}}

\def\eps{\epsilon}

\newcommand{\vev}[1]{\langle#1\rangle}
\def\be{\begin{equation}}
\def\ee{\end{equation}}
\newcommand{\eref}[1]{(\ref{#1})}
\newcommand{\Eref}[1]{Eq.~(\ref{#1})}
\newcommand{\rem}[1]{}

\def\half{{1\over 2}}
\def\NN{{\cal N}}

\def\none{$\NN=1$}

\def\susic{supersymmetric}

\def\OO{{\cal O}}
\def\ZZ{{\bf Z}}

\def\tim{\tilde m^2}
\newcommand{\dbyd}[2]{{\partial #1\over\partial #2}}

\def\ltap{\ \raise.3ex\hbox{$<$\kern-.75em\lower1ex\hbox{$\sim$}}\ }
\def\gtap{\ \raise.3ex\hbox{$>$\kern-.75em\lower1ex\hbox{$\sim$}}\ }

\title{Suppressing Flavor Anarchy}
\author{Ann E. Nelson\\ 
Department of Physics, Box 1560, University of Washington,\\
           Seattle, WA 98195-1560, USA}
\author{Matthew J. Strassler\\
School of Natural Sciences, Institute for Advanced Study,\\
Princeton, NJ 08540, USA }
\preprint{UW/PT-00/01\\IASSNS--HEP--00/46}

\abstract{ We present a new mechanism, which does not require any
flavor symmetry, to explain the small Yukawa couplings and CKM mixing
angles.  The Yukawa matrices are assumed to be random at short
distances and the hierarchical structure is generated in the infrared
by renormalization group flow. The generic qualitative predictions of
this mechanism are in good agreement with observation. We give several
simple examples in supersymmetric theories. We show that our mechanism
can also ameliorate the supersymmetric flavor problem, and make
predictions for the superpartner mass spectrum.  The mechanism is
fully consistent with grand unification, and in $SU(5)$-based models
of neutrino mass, predicts a large mixing angle for
$\nu_\mu\leftrightarrow\nu_\tau$ oscillations. }

\begin{document}
%
%

\section{Introduction}

\label{intro}
The quark and lepton masses and mixing angles exhibit a degree of
structure and regularity which is puzzling, as in the Standard Model
and typical extensions these are determined by arbitrary Yukawa
coupling matrices. In particular, the three generations of quarks and
leptons have quite disparate masses spanning more than five orders of
magnitude, with rather small mixing angles between the different
generations, and with the largest mixing between adjacent
generations. 

In this paper we show how such a pattern can arise as the result of
renormalization group running, when some of the quarks and leptons
acquire substantial anomalous dimensions that persist over a large
range of energy scales.  We give supersymmetric examples where such
anomalous dimensions are the result of superpotential couplings of the
light quarks and leptons to a superconformal sector.  The sizes of the
resulting Yukawa couplings are set by an accidental and approximate
$U(1)_R$ symmetry, whose presence is a consequence of the approximate
superconformal symmetry in the strong-coupling sector.  The resulting
predictions resemble those of successful flavor models with horizontal
abelian symmetries and with all standard model fermions carrying
charges with the same sign
\cite{Froggatt:1979nt,Leurer:1993wg,Leurer:1994gy,Grossman:1995hk}.
In our scenario, the R-symmetry which sets the charges arises
dynamically and accidentally, and unitarity demands all matter fields
have positive charges; thus these features do not have to be imposed
by hand.

We also examine the soft supersymmetry breaking terms in these models.
The trilinear scalar interactions associated with superpartners of the
light fermions are dramatically reduced, and have similar texture to
the Yukawa matrices. They are not, however, necessarily proportional
to Yukawa couplings, leading to interesting flavor violating
signatures.  In a large class of models, the renormalized masses of
the scalar superpartners of the light fermions become nearly
degenerate, suppressing the supersymmetric contribution to
flavor-changing processes.

\section{How a CFT can produce the flavor hierarchy}
\label{cft}
Typically, dimensionless coupling constants such as Yukawa couplings
have logarithmic energy dependence, and so renormalization group flow
cannot account for very large ratios of dimensionless couplings, even
in a strongly-coupled theory.  This is because the effective theory is
describable in terms of weakly-coupled degrees of freedom at most
energies and is only truly strongly coupled near one particular energy
scale (although the weakly-coupled degrees of freedom may be different
in the ultraviolet and infrared).  However it was realized some time
ago \cite{Georgi:1983mq} that in a theory with a nearly scale
invariant gauge coupling the Yukawa couplings can run like a power of
the renormalization group scale $\mu$, and that a large Yukawa
hierarchy could result from renormalization group running.

Many field theories which are free in the ultraviolet will flow to a
Conformal Fixed Point (CFT) in the infrared, such as the simple
example discussed in ref.~\cite{Banks:1982nn}.  A prototypical
supersymmetric example is supersymmetric QCD with $\frac32
N_c<N_f<3N_c$ \cite{Seiberg:1995pq}.  In this section, we examine the
effect of coupling the standard model to an approximate supersymmetric
CFT.  At the CFT, the operators of the free theory develop anomalous
dimensions of order one. In many supersymmetric cases, these anomalous
dimensions are computable even at strong coupling, using
superconformal symmetry. The large anomalous dimensions cause rapid
running of the Yukawa couplings.  Such anomalous Yukawa coupling
scaling leads to a simple explanation of the hierarchy of quark and
lepton Yukawa couplings. As we will show in section \ref{generic}, the
generic predictions of this mechanism, resembling the most successful
models of
\cite{Froggatt:1979nt,Leurer:1993wg,Leurer:1994gy,Grossman:1995hk},
give qualitative relationships between mass ratios and mixing angles
which are in striking agreement with reality.

\subsection{Effect of a Conformal Field Theory on a Weakly Coupled Sector}

The effect of coupling a superfield $X$ of the MSSM to a CFT via the
superpotential was considered in Sec. (2.1)  of \cite{Nelson:1997km}.  We
summarize the results of that section here.

Consider an \none\ \susic\ gauge theory with gauge group $G$, charged
matter $Q$, neutral matter $X$, and a superpotential $W(Q,X)$.
Suppose that this theory becomes conformal in the infrared.  Unitarity
then requires that all gauge-invariant operators $\OO$ (except the
identity operator) have dimension $\dim\OO$ greater than or equal to
one; if $\dim\OO=1$, then the operator satisfies the Klein-Gordon
equation $\partial^2\OO=0$ \cite{Mack:1977je,Flato:1984te}.   By assumption
$Q$ is not gauge invariant, although there will be gauge-invariant
chiral operators $\OO(Q)$ which are multilinear in $Q$.  This means
$Q$ may have a negative anomalous dimension.  By contrast, $X$ is
gauge invariant, so $\dim X\geq 1$ and its anomalous dimension
$\gamma_X = 2[\dim X-1]\geq 0$; if $\gamma_X=0$ then $X$ must decouple
in the infrared from the remainder of the theory.

What is the effect of this positive anomalous dimension?  First,
recall that at any superconformal fixed point there is an anomaly-free
$U(1)$ R-symmetry; the dimension of a chiral operator $\OO$ is
determined by its charge under the R-symmetry, $\dim(\OO) = {3\over 2}
R_\OO$. It follows that if $\OO$ and $\OO'$ are chiral then
$\dim(\OO\OO')= \dim\OO+\dim\OO'$.  Now let us consider
a given term $cw$ in the superpotential $W$, where $w$ is an operator
and $c$ is its coefficient.  Note $\dim c + \dim w = \dim W = 3$.
The operator $w$ must contain $k\geq 0$ powers of neutral fields $X$
times a $G$-invariant chiral operator $\OO$ built from the $Q$ fields.
If $\OO$ is not the unit operator, then $\dim\OO\geq 1$, so $\dim w
\geq k+1$; similarly, if $\OO$ is the unit operator, $\dim w \geq k$.

The beta function for 
$c$ is simply
$$
\dbyd{c}{\ln\mu} \equiv \beta_c = c (\dim w-3)
$$
so if $\dim w > 3$ in the infrared, $c$ will flow to zero and the term
$cw$ will disappear from the superpotential of the
CFT. \footnote{The possibility that relevant or marginal operators
in the K\"ahler potential couple the fields $X$ to the CFT is not
excluded by this argument.  However, no examples of this phenomenon
are known. We will assume such couplings do not occur in
the CFTs that we discuss below.}
Therefore, at or near such a conformal fixed point, almost all operators
that can appear in the superpotential are irrelevant.  The only
marginal or relevant operators are of the form $\OO(Q)$ (for
$\dim\OO\leq 3$,) $X\OO(Q)$, (for $\dim\OO$ strictly less than $2$,)
or $XX$ (if $\dim XX\leq 3$.)  Note that if $X\OO(Q)$ is to appear in
the superpotential of the CFT, the constraint $1<\dim \OO < 2$
generally implies that $0<\gamma_X<2$. (This last constraint can be
avoided if $\OO$ is redundant, that is, set equal to a total
derivative by the equations of motion.  In this case  $\OO$ is not
primary and the relation between R-charge and dimension is modified.
However, the R-charge of $\OO$ still determines the R-charge, and
thereby the dimension, of $X$.)

In particular, consider what happens to a Yukawa coupling of the form
$yX_1X_2X_3$, where the $X_i$ are all neutral under $G$.  This has has beta
function 
$$
\beta_y = {y\over 2}(\gamma_1+\gamma_2+\gamma_3)
$$
where $\gamma_i$ is the anomalous mass dimension of the field $X_i$ at
or near the conformal fixed point of $G$.  As emphasized above, these
anomalous dimensions must be positive, and so $y$ flows to smaller
values in the infrared:
\be\label{yrun}
y(\mu) = y(\mu_0)\times
\left({\mu\over\mu_0}\right)^{\half(\gamma_1+\gamma_2+\gamma_3)}
= y(\mu_0)\times
\left({\mu\over\mu_0}\right)^{\dim(X_1X_2X_3)-3}\ .
\ee 

Now suppose that the CFT in question has an exact $SU(3)\times
SU(2)\times U(1)$ global symmetry, under which some of the $X$ and $Q$
fields are charged.  If we weakly gauge this symmetry, with the
standard model gauge couplings $g_i$, the effect on the CFT is small;
both beta functions and anomalous dimensions change by of order
$g_i^2/16\pi^2$.  If the anomalous dimensions of the $X$ fields were
already large for $g_i=0$, then the physics of the approximately
conformal theory with non-zero $g_i$ is qualitatively and
quantitatively similar to that of the CFT with $g_i=0$.

Thus, we consider the following scenario. In the effective theory
below the scale of quantum gravity, we have  gauge group $G\times
SU(3)\times SU(2)\times U(1)$, with $G$ strongly coupled and the
standard model relatively weakly coupled.  If the quarks and leptons
are coupled in the superpotential to the charged matter of $G$,
conformal dynamics associated with $G$ will cause them to develop
large anomalous dimensions.  These in turn will cause their Yukawa
couplings to the Higgs bosons to run to small values, as in \Eref{yrun}.

More specifically, consider for example the leptons and their Yukawa
couplings $y_{ai}L_a H_d E_i$, where $L_a$ and $E_i$ ($i,a=1,2,3$) are
the doublet and singlet leptons of the three generations, and $H_d$ is
the down-type Higgs doublet.  Suppose a CFT caused $L_a$ and $E_i$ to
develop anomalous dimensions $\gamma_a$ and $\gamma_i$ respectively.
Let us imagine that all Yukawa couplings are of order one at the
Planck or string scale $M_0$.  Let us further imagine that the gauge
coupling of $G$ and the Yukawa couplings involving $Q$ become
conformal near $M_0$, and that the CFT decouples from the standard
model (through dynamics we will discuss later) at a scale
$\mu=M_c$. Define
\be
\eps^L_a \equiv \left({M_c\over M_0}\right)^{\half\gamma_a} \ ; \
\eps^{\bar E}_i \equiv \left({M_c\over M_0}\right)^{\half\gamma_i}  \ .
\ee
Then, as in \eref{yrun}, the Yukawa couplings will run down to values
\be
y_{ai}(M_c) \sim \eps^L_a \eps^{\bar E}_i\ .
\ee
Below this scale the usual RG equations of the standard model will
apply.  Thus, if the $\gamma_a,\gamma_i$ have a hierarchy of order
one, this hierarchy is exponentiated into a large lepton mass
hierarchy.  Since we want to obtain an electron Yukawa coupling of
order $3\times 10^{-6}$, and since $\gamma_e$ typically cannot be larger
than 2, we will typically need $M_0/M_c \sim 10^2 - 10^8$,
depending on the details of the model.  Note this can be much less
than $M_0/M_W$, and can even be of order $M_0/M_{GUT}$ in some cases.

The lesson from the above paragraph is simple.  Each quark or lepton
$X$ coupled to the fields $Q$ in the superpotential will develop a
large anomalous dimension $\gamma_X$, and an associated suppression
factor
$$
\eps \equiv \left({M_c\over M_0}\right)^{\half\gamma_X}  = 
\sqrt{Z_X(M_0)\over Z_X(M_c)}\ ,
$$
where $Z_X(\mu)$ is the Wilsonian wave-function renormalization factor
for $X$ at the scale $\mu$. This means that a Yukawa coupling
$y_{12}X_1X_2H$, where $H$ is a Higgs boson, will run down to
\beq
\label{eq:eps}
y_{12}(M_c) \sim y_{12}(M_0)\eps_1\eps_2 \ .
\eeq
where $\eps_i$ is the suppression factor for $X_i$.

Although this scenario does not lead to precise predictions, it does
give order-of-magnitude relations.  There are three possible sources
of such relations.  First, all supersymmetric CFTs have a special
$U(1)$-R symmetry which determines anomalous dimensions for all chiral
operators $\OO$, through the relation $\dim \OO = {3\over 2}R_\OO$ 
\cite{Flato:1984te,Dobrev:1985qv}.
When (as is not always the case) this symmetry can be uniquely
determined, $\gamma_X$ can be calculated (up to corrections of order
standard model loop-factors) and so $\eps_X$ is known for all quarks
and leptons as a function of $M_c/M_0$.  Even if $M_c$ is a free
parameter, there are still predictions, since when the $\gamma_X$ are
known, there will be mass ratios in which $M_c$ cancels out.

Second, even if $\gamma_X$ is not known, it may be that there are
GUT-type or other relations between the $\eps$ suppression factors for
different standard model multiplets which reduces the number of
parameters.  For example,  
if an $SU(5)$ relation links the suppression factors of
doublet quarks and up-type antiquarks, it leads to 
order-of-magnitude predictions for the mixing angles of the CKM
matrix.  
Similar $SU(5)$ relations lead to predictions for lepton masses and
neutrino oscillation parameters, as we will discuss in section \ref{sufive}.

Finally, even if all the $\eps$'s are left as free parameters, there
are still some generic predictions for quark and lepton mixing angles,
as will be discussed in greater detail in section \ref{generic}.  We
will see that all such predictions either are satisfied in nature or
have not yet been tested.

Note that to obtain large suppression factors does not require an
exact CFT.  A near-CFT  with slow drift, or a flow from one
approximate CFT to another, will imply anomalous dimensions for quarks
and leptons which are not constants.  This will make it difficult to
compute the suppression factors $\eps_i$.  However, this is a problem
only for model builders and experimental verification, not one for
nature.  Obtaining a reasonable fermion mass spectrum only requires
that anomalous dimensions be large, but not necessarily constant, over
a significant range of energies. 

\subsection{Comments on the Origin of the Generational Hierarchy} 

The reader may be concerned that we will need to impose flavor
symmetries to ensure the generational hierarchy, and particularly to
assure that the top quark does not develop a substantial anomalous
dimension, which would drive its Yukawa coupling unacceptably small.
Our scenario makes no such requirement.  In particular, the model at
the Planck scale has no notions of generational structure; all Yukawa
couplings are of order one, and thus all mixings are large.
Generational structure is an {\it output} of the model, not an input.

An essential property of \none\ supersymmetric conformal field
theories is that they always contain an anomaly-free $U(1)$
R-symmetry, under which all gauge-invariant operators carry a definite
charge.  This symmetry need not be present away from the conformal
point; it may be violated by a host of irrelevant operators, and may
exist only as an accidental symmetry in the infrared.  At an
approximate fixed point it of course will be  only approximately
conserved. 

In the scenario discussed here, we have an approximate conformal fixed
point, whose conformal invariance is weakly broken by the standard
model gauge and Yukawa couplings.  We therefore have an approximate
$U(1)$ R-symmetry that the couplings of the model associated with
 the strong dynamics --- those which would still be
present if the conformal invariance were exact --- must respect. Note
that we need not impose that our theory have a $U(1)$ R-symmetry in
the ultraviolet; the dynamics will ensure that the theory develops an
accidental approximate $U(1)$ in the regime where the theory is
approximately conformal.

The existence of this $U(1)$ is essential to our mechanism.  The
standard model matter superfields are gauge-invariant operators with
respect to the conformal sector.  Consider, say, the three generations
of lepton doublets $L^{(0)}_i$.  In an exactly conformal regime, there
must exist three orthogonal linear combinations $L_i$ of the
$L^{(0)}_i$ which carry definite and generally distinct R-charge
assignments.  {\it Consequently, these three combinations cannot mix
with one another and have distinct anomalous dimensions.}  This
ensures that each standard model multiplet can be taken to have its
own suppression factor.  These statements remain approximately true
when the conformal invariance is only approximate.

As illustration, suppose an exact CFT has operators $\OO_a$ with
R-charge $r_a$ and dimensions $d_a=\frac32 r_a$.  A standard model
field $X_i$ with R-charge $r_i$ can couple only to those operators
with $r_a+r_i=2$.  More precisely, if a term $X_i \OO_a$ appears in
the superpotential of a CFT, then the term $X_i\OO_b$ cannot appear if
$r_a\neq r_b$.  Thus, suppose the operators with couplings to the
lepton doublets $L^{(0)}_i$ have R-charges $r_a$ with
$r_1<r_2<r_3<r_4<\dots$.  In this case, only one linear combination of
the $L^{(0)}_i$ can couple to $O_1$.  We may take this by definition
to be $L_1$.  In a similar way we may define $L_2$ to be the linear
combination which couples to $\OO_2$, and $L_3$ to be that which
couples to $\OO_3$.  No linear combination will couple to $\OO_4$
(unless a nongeneric cancellation has made one of the previous
couplings zero.)  Since $L_1$, $L_2$ and $L_3$ in this basis have
different R-charges, they are orthogonal and do not mix; their kinetic
terms are diagonal. 

We may now ask how the theory approaches this fixed point, starting
from an arbitrary K\"ahler potential and superpotential.  The answer
is that the initial condition doesn't matter, as long as it is
reasonably generic.  Whatever the starting point, the approach to the
CFT will ensure that the low-energy theory can be written conveniently
in a basis with diagonal kinetic terms and diagonal $X_i\OO_a$
couplings in the superpotential.  In this basis, each of the $X_i$
will have a definite anomalous dimension and a corresponding
suppression factor $\epsilon_i$.  Thus, although mixing between the
$L_i$ will not be strictly zero in the approximately conformal regime,
it will sufficiently reduced that the key approximation --- that each
field has its own suppression factor --- will be a good
one.\footnote{Let us forestall a possible confusion.  The natural
basis in supersymmetric theories puts all of the renormalization into
the wave function, leaving the superpotential unrenormalized. In this
case one may well ask how a non-diagonal coupling matrix in the
superpotential can evolve into a diagonal one!  However, the {\it
physical} coupling of the $X_i$ to the $\OO_a$ is not determined by
the superpotential, but is given by dressing the superpotential by the
wave function matrices of $X_i$ and $\OO_a$.  These matrices can
diagonalize the superpotential couplings, so that the physical
interactions respect the $U(1)$ R-symmetry.}

For the $L_i$ fields in the situation just discussed, all three will
have positive anomalous dimensions and all three will therefore have
suppressed Yukawa couplings.  However, suppose that among the
operators $\OO_a$ with charges conjugate to those of isospin-doublet
quarks, only $\OO_1$ and $\OO_2$ have dimensions less than 2.  In this
case only the couplings $Q_1\OO_1$ and $Q_2\OO_2$ can appear in the
CFT; the orthogonal linear combination $Q_3$ will decouple, and will
have no anomalous dimension in the CFT regime.  The field $\bar U_3$
may similarly decouple if only two operators with appropriate charge
have dimension less than 2.  If both of these conditions hold, then
the Yukawa coupling for $Q_3\bar U_3$ will be unsuppressed, and the
top quark will be heavy.\footnote{Since the physical value of $y_t$
is $\sim 1$, not $4\pi$ or $\sqrt{4\pi}$, there is room for $y_t$ to
be reduced a small amount from its value at $M_0$.  Approximate,
rather than strict, decoupling of the top quark from the strong sector
is therefore sufficient to maintain a reasonable mass prediction.}
Thus, to obtain a heavy top quark while obtaining the remaining
hierarchy of the standard model merely requires choosing a CFT with an
appropriate spectrum of operators $\OO_a$.  The top quark need not be
labelled in advance; the near-conformal dynamics will determine it for
us.

Note that this argument would fail if the strong dynamics never
becomes approximately conformal.  In this case, no linear combination
of the $Q_i$ would decouple cleanly from the CFT; the $Q_i$ would
continue to mix.  A similar fate would befall the $\bar U_i$.  In this
case the up-type Yukawa matrix might not have a large eigenvalue, nor
would intergenerational mixing be naturally suppressed.

\section{Generic Predictions for Masses and Mixing Angles } 
\label{generic}

The main consequence of this mechanism is that 
low energy Yukawa coupling matrices are
of the form
\beq 
\label{eq:qual}
Y_{ij}={{\eps}_L}_i{{\eps}_R}_jY^{(0)}_{ij}\ .
\eeq
where $Y^{(0)}_{ij}$ is the short distance matrix. 
 Assuming
generic  $Y^{(0)}_{ij}$,  the qualitative prediction of 
\Eref{eq:qual} is that when the $\eps$ factors are different for
different
fields, then  
 the mass matrices are approximately diagonal.  These matrices are
diagonalized by
left- and right-handed mixing matrices with  mixing angles
 between the $i^{th}$ and $j^{th}$ generations ${\theta_{L,R}}_{ij}$,
of order
\beq
\label{eq:genang}
{\theta_L}_{ij}\sim {{{\eps}_L}_i\over{{\eps}_L}_j}
\sim {\theta_L}_{ji},\quad 
{\theta_R}_{ij}\sim {{{\eps}_R}_i\over{{\eps}_R}_j}
\sim {\theta_R}_{ji},\quad \ (i<j) .
\eeq 
Also, the $m_i$ satisfy the order of magnitude relations 
\beq
\label{eq:genmass}
{m_i\over m_j}\sim{{{\eps}_L}_i{{\eps}_R}_i\over {{\eps}_L}_j{{\eps}_R}_j} \ .
\eeq

Yukawa matrices with the  structure of eq.~\eref{eq:qual} have
appeared previously in Froggatt-Nielsen models \cite{Froggatt:1979nt}
and their generic consequences are well known
\cite{Froggatt:1979nt,Leurer:1993wg,Leurer:1994gy,Grossman:1995hk}. 
In the rest of this section we summarize these predictions.

In the standard model, only the left-handed mixing angles are
observable, and so we can not verify all the predictions of
eqs.~\eref{eq:genang} and \eref{eq:genmass} without making additional
assumptions, such as grand unification.  However the generic
predictions of eq.~\ref{eq:genang} for the CKM matrix are 
\beq
V_{ud}&\sim& V_{cs}\sim V_{tb}\sim \CO(1)\\
V_{ub}&\sim &V_{td}\sim
V_{cb}V_{us}\\
V_{ts}&\sim&V_{cb}\\ V_{us}&\sim&V_{cd}\ , 
\eeq 
all of which are satisfied to within a factor of
two.  The last two equations follow from the first two
and unitarity (if there are only three generations.)
Note generational ordering is automatic: the CKM matrix is most
nearly diagonal in the basis where both the up and the down-type
quarks are ordered by mass.

Similarly, for neutrinos, if there is
a basis in which the mixing matrix is nearly diagonal, then the lightest
neutrino is mostly of the electron type and the next lightest of the muon
type. In the neutrino sector we can  also predict (not very stringent)
order-of-magnitude bounds on $V^L$, the lepton analogue of the CKM matrix: 
\beq \label{nuangle}
V^L_{e1}&\sim& V^L_{\mu 2}\sim V^L_{\tau3}\sim \CO(1)\\
V^L_{e2}&\sim &V^L_{\mu1}\\
V^L_{\mu3}&\sim& V^L_{\tau 2}\\
V^L_{e3}&\sim &V^L_{\tau 1}\sim V^L_{e2}V^L_{\tau 2} \label{nuangleD} \ .
\eeq
These remain to be verified, but are allowed.  

Other generic
predictions of \Eref{eq:genmass} are lower bounds on the order of magnitude
of left-handed mixing angles in terms of mass ratios. These lower bounds are
saturated when the right-handed $\eps$'s are all assumed to be the same.  For
the CKM matrix these lower bounds are all  satisfied. 
\def\agt{\gtap} 
\beq
V_{us}&\agt&{\rm Max}\left({m_d\over m_s},{m_u\over m_c}\right)\\
V_{cb}&\agt&{\rm Max}\left({m_s\over m_b},{m_c\over m_t}\right)\ .
\eeq
In the lepton sector we do not know all the mixing angles or the mass ratios,
but we may predict 
\beq \label{nuangleE}
 V^L_{e2}&\agt &
{\rm Max}\left({m_e\over m_\mu},{m_{\nu_1}\over m_{\nu_2}}\right)\\
V^L_{\mu3}& \agt &
{\rm Max}\left({m_\mu\over m_\tau}, {m_{\nu_2}\over m_{\nu_3}} \right)
\label{nuangleF}
\eeq
which are  also allowed.

We may make  more predictions for the neutrino oscillation parameters by
making a reasonable assumption about  the right-handed
neutrinos.  We assume there are at least three of these, which obtain large
Majorana masses at a common scale $M$.   If  the only hierarchy in the
right handed neutrino masses is due to    suppression factors, then these will 
appear in  both the right-handed neutrino Majorana mass matrix and in the
neutrino couplings to the Higgs.  After diagonalizing the neutrino mass matrix,
there are three light eigenstates,  which are essentially left-handed weak
doublet neutrinos with small Majorana masses, and mass matrix of the form
\beq
{M_\nu}_{ij}\sim {\eps_L}_i{\eps_L}_j\lambda^{(0)}_{ia}\lambda^{(0)}_{jb}
{\eps_R}_a{\eps_R}_b ({\eps_R}_a{\eps_R}_b M^{(0)}_{ab})^{-1}\ .
\eeq
Note that the right-handed neutrino suppression factors ${\eps_R}$ cancel out
in the above expression and do not affect low energy phenomenology.
We can therefore predict the order of magnitude relations
\beq \label{numass}
\sqrt{m_1\over m_2}&\sim&{V^L_{e2}}\\
\sqrt{m_2\over m_3}&\sim&{V^L_{\mu3}} \ .
\eeq
These predictions are not consistent with the ``just-so'' solution 
\cite{Barger:1981xs,Glashow:1987jj,Acker:1991zj} to the solar
neutrino problem \cite{Haxton:1995hv,Bahcall:1995bt}, 
unless all the neutrino masses are of order $10^{-5}$ eV,
which is inconsistent with the superKamionkande evidence for large angle
$\nu_\mu\leftrightarrow\nu_x$ neutrino
oscillation\cite{Fukuda:1998mi} with a mass squared difference of
order $10^{-2}$---$10^{-3}$~eV$^2$.  
With three light
Majorana neutrinos, the latter must be interpreted as nearly maximal
$\nu_\mu\leftrightarrow\nu_\tau$ oscillation, and so 
${\epsilon_L}_\mu\sim {\epsilon_L}_\tau$. 
If one wishes to solve  the solar  neutrino problem via the
MSW effect \cite{Wolfenstein:1978ue,Mikheev:1985gs,Mikheev:1986wj},
then one needs 
\cite{Fogli:1994ck,Fogli:1996ne,Hata:1997di,Bahcall:1999ed,Bahcall:1998jt,Fogli:1999zg}
\beq\label{anarchy}
{m_2^2\over m_3^2}\sim  10^{-2}\ .
\eeq  
A mass ratio $m_2/m_3\sim 1/3$---$1/10$ can easily result in a seesaw
model from random neutrino mass matrices; see for example
ref.~\cite{Hall:1999sn}.  Thus eq. \eref{anarchy} does not require any
hierarchy between $\eps_2$ and $\eps_3$, and is consistent with
attributing the superKamiokande result to large angle
$\nu_\mu\rightarrow\nu_\tau$ oscillations. If $\eps_2\sim\eps_3$ then
the $V^L_{e2}$ and $V^L_{e3}$ mixing matrix elements should be of
comparable magnitude. The small-angle MSW solution
\cite{Bahcall:1999ed} is consistent with this, as is the CHOOZ
experiment \cite{Apollonio:1997xe}, if we take both angles of order
$.04$.  By contrast, the large angle MSW solution
\cite{Bahcall:1999ed} is in greater difficulty, since it would require
$m_3^2-m_1^2\sim m_3^2- m_2^2 \gg m_2^2-m_1^2$ and $ V^L_{e2} \sim 1$,
which is difficult to reconcile with the superKamiokande and CHOOZ
experiments \cite{Apollonio:1997xe} taken together unless
$V^L_{e3}\ltap 0.15$.  However, as shown in ref.~\cite{Hall:1999sn},
there is actually a reasonable amount of parameter space in which an
anarchic neutrino mass matrix happens to give a sufficiently small
$V_{e3}$ and large $V_{e2}$ and $V_{\mu3}$.  We therefore do not view
this scenario as clearly inconsistent with the large-angle MSW
solution.

 It is possible to obtain a neutrino mass hierarchy which is
substantially {\it larger} than eq.~(\ref{numass}).  Suppose the scale
at which the Majorana masses are generated is not unique, so that a
hierarchy in the right-handed neutrino masses is present which is not
associated with the $\eps$ suppression factors.  This structure will
generate a hierarchy of the low-energy left-handed neutrino masses
which is not given by the suppression factors.  However, provided that
all couplings of the {\it left}-handed neutrinos are proportional to
their $\eps$ suppression factors, then the mixing angles, and thus
predictions eqs.~(\ref{nuangle})--(\ref{nuangleD}) and
(\ref{nuangleE})--(\ref{nuangleF}), will not be affected.

\section{Generic $SU(5)$ based Predictions for Masses and Mixings}
\label{sufive}
Current available data on mass and mixing parameters hint strongly at
$SU(5)$, either as a grand-unification symmetry or as an approximate
symmetry of strong flavor dynamics.  This has been pointed out in the
context of ``ten-centered'' models
\cite{Babu:1996hr,Strassler:1996ia} and other flavor models
\cite{Albright:1998vf,Altarelli:1998ns}.  In ten-centered  models, a universal
suppression factor is assumed for the three $\bar 5$'s of $SU(5)$, and
the generational hierarchy is entirely due to different suppression
factors for the different $10$'s.  These models have only four
parameters, which they use to predict the nine charged fermion masses
and the CKM angles.  Most of the predictions are successful; two out
of eight (the Cabibbo angle and the electron-to-muon mass ratio) are
off by factors of $5$ and $1/8$ respectively
\cite{Babu:1996hr,Strassler:1996ia}.  A single additional parameter
predicts the properties of the neutrino masses and mixings; in
particular, large angle mixing and a small hierarchy are natural and
generic in  these models \cite{Babu:1996hr,Strassler:1996ia,Strassler:Nagoya}.

Here, we take a somewhat wider view, increasing the number of
parameters by two beyond the ten-centered case, but obtaining a better
fit.  In particular, we allow for independent suppression 
factors $\eps_{10_i}$ and $\eps_{\bar 5_r}$.  Rather than assume a particular CFT, we
use data to fit for the best values of the $\eps$'s, and
use these values to make order-of-magnitude predictions in the standard
model. The predictions obtained are    
successful.  This wider framework can more easily
accommodate the small  mixing angle solution to the solar neutrino
problem than can the ten-centered models. It is encouraging that 
when one fits
  the suppression factors from the quark and lepton masses and CKM matrix, 
one obtains similar suppression factors for the left-handed 
$\mu$ and $\tau$ doublets. Thus $SU(5)$ symmetry
predicts unsuppressed mixing angle for
$\nu_\mu\leftrightarrow\nu_\tau$ oscillations. 

The    predictions of $SU(5)$ based
models
are (in addition to the generic predictions of the previous section):
\beq\label{sufiveeq}
\sqrt{m_c/m_t}&\sim & V_{cb}\sim V_{ts}\sim\eps_{10_2}/\eps_{10_3}\\
\sqrt{m_u/m_c}&\sim & V_{us}\sim V_{cd}\sim\eps_{10_1}/\eps_{10_2}\\
{m_\mu/m_\tau\over\sqrt{m_c/m_t}}&\sim&
{m_s/m_b\over\sqrt{m_c/m_t}}\sim V_{\mu3}\sim\sqrt{m_{\nu_2}/m_{\nu_3}}
\sim\eps_{\bar 5_2}/\eps_{\bar 5_3}\label{mutau}\\
{m_e/m_\mu\over\sqrt{m_u/m_c}}&\sim&
{m_d/m_s\over\sqrt{m_u/m_c}}\sim V_{e2}\sim \sqrt{m_{\nu_1}/m_{\nu_2}}
\sim\eps_{\bar 5_1}/\eps_{\bar 5_2}\ .\label{emu}
\eeq
Here all masses are to be assumed to be renormalized at a common, high
scale.

Of course the suppression factors are somewhat uncertain, since the
low-energy spectrum results from a combination of the suppression
factors with numbers of order one from the assumed high-energy
anarchy.  Eq.~(\ref{emu}) is especially ambiguous, since $m_e/m_\mu$
is smaller than $m_d/m_s$ by about a factor of 10. If we take the
geometric mean of the various determinations of mass ratios and mixing
angles which go into determining the suppression factors we get the
estimates:
\beq\label{epsilons}
\eps_{10_2}/\eps_{10_3}&\sim&\sqrt{V_{cb}\sqrt{m_c/m_t}}\sim 0.04 \\
\eps_{10_1}/\eps_{10_2}&\sim&\sqrt{V_{us}\sqrt{m_u/m_c}}\sim 0.07\\
\eps_{\bar 5_2}/\eps_{\bar 5_3}&\sim
&{\sqrt{(m_s/m_b)(m_\mu/m_\tau)}\over \eps_{10_2}/\eps_{10_3}}\sim 0.9
\\ \label{mue}
\eps_{\bar 5_1}/\eps_{\bar 5_2}&\sim 
&{\sqrt{(m_d/m_s)(m_e/m_\mu)}\over
\eps_{10_1}/\eps_{10_2}}\sim 0.15\ .
\eeq
In these expressions we have used numbers more appropriate for
the GUT scale than for the weak scale, especially for $m_t$.
    
The leptonic weak mixing angles depend on the $\eps_{\bar 5}$'s and
are predicted to be 
\beq\label{neutrino}
V_{e2}&\sim&\eps_{\bar 5_1}/\eps_{\bar 5_2}
\sim 0.15\\
V_{\mu3}&\sim&\eps_{\bar 5_2}/\eps_{\bar 5_3}
\sim 0.9
\eeq
which is consistent with essentially maximal $\nu_\mu\leftrightarrow
\nu_\tau$ oscillations and, through eq~(\ref{mutau}), a relatively
small mass hierarchy.  Eq.~\eref{neutrino} lies  between the
expectation for either the small mixing angle or the large mixing
angle MSW solution to the solar neutrino problem.  This is
unfortunate, since given the uncertainties from the assumed anarchy in
the neutrino matrices, we cannot say that either is ruled out,
or even that one is significantly favored.

\section{Additional features: Nucleon decay and the Flavor Problem}
\label{decay}

Another advantage of this scenario is that it suppresses nucleon decay
to acceptable levels.  The mechanism is analogous to the one typically
observed in compositeness models, as noted in
\cite{Nelson:1997km,Strassler:1996ia,Strassler:Nagoya,Babu:tenc}.

Even if R-parity is used to avoid dimension-four operators (here the
dimension-counting refers to operators in the Lagrangian, not the
superpotential) which could cause rapid nucleon decay, dimension-five
operators from superpotential terms of the form $Q_iQ_jQ_kL_r$ and
$\bar U_i\bar U_j\bar D_r \bar E_k$ will lead to rapid nucleon decay
 if these operators are suppressed only by the natural coefficient
$1/M_{pl}$.  In particular, as shown in \cite{Hisano:1993jj}, the
coefficients of these operators for first-generation fields cannot
exceed $(10^{-6}$---$10^{-7}) M_{pl}^{-1}$ when first generation
particles are involved.

Fortunately, in our scenario each field comes with a suppression
factor related to its mass.  It turns out that for moderate values of
$\tan \beta$ (the ratio of the expectation values of the up-type and
down-type Higgs bosons) the coefficients are sufficiently small. The
decay mode $p\to K^+\nu$ is typically dominant.  For small $\tan\beta$
we expect \cite{Babu:1998ep,Babu:tenc} the largest contribution to come from
the $Q_2Q_2Q_1L_r$ operator, along with a Cabibbo angle mixing.  This
gives a coefficient of order
$$
\eps_{10_2}^2\eps_{10_1}V_{us}\eps_{\bar 5_r}\sim
3\cdot 10^{-6}\cdot\eps_{10_3}^3\eps_{\bar 5_r}
$$  
which requires a suppression from $\eps_{10_3}$ and/or $\eps_{\bar 5_r}$.
Note that $\eps_{10_3}$ could be as small as, say, $1/3$, while
$\eps_{5_3}$ could be as small as $(\tan\beta/60)\eps_{10_3}$.  Thus,
sufficient suppression is certainly possible, although discovery 
should be feasible.

However, at large $\tan\beta$ the operator $\bar U_1\bar D_r\bar
U_3\bar E_3$ typically gives the dominant contribution
\cite{Babu:1998ep,gotonihei,Babu:tenc}.  This occurs through Higgsino exchange
which converts the operator $ud\tilde t\tilde\tau$ to $uds\nu_\tau$.
The operator is proportional to $\eps_{10_3}^2\eps_{10_1}\eps_{\bar 5_r}$,
and the Higgsino exchange gives
the Yukawa couplings $y_t y_\tau$ and a mixing angle $V_{ts}$.
Altogether this gives an effective coupling of order 
$$
y_ty_\tau V_{ts}\eps_{10_3}^2\eps_{10_1}\eps_{\bar 5_r}
\sim 
10^{-4}{\tan\beta\over 60}\eps_{10_3}^3\eps_{\bar 5_r}
$$
This dominates the previous process for large $\tan\beta$.
The exact value of $\tan\beta$ where this operator produces excessive
proton decay   depends on many unknown parameters, such as
superpartner masses, fermion
mixing angles,  and hadronic uncertainties. However it seems unlikely
that $\tan \beta$ of order 50, as would be needed to explain $m_b/m_t$
without suppression factors for the right handed bottom quark, can be
made consistent with acceptably small proton decay. This
problem is not limited to our scenario \cite{Babu:1998ep,Babu:tenc}; many
grand unified and more general flavor scenarios are in serious trouble
at large $\tan\beta$ due to the absence of proton decay.  (Of course
particular models may escape these generic constraints \cite{escape}.)

It is also generic in these models that charged-lepton branching
fractions are not negligible at small $\tan\beta$
\cite{Babu:1998ep}. In particular, the special
cancellations which occur in minimal GUTs are generally absent in
these models.  Since the muons and electrons which may appear come
from the fields $L_2$ and $L_1$, we expect the ratios of their
branching fractions to be of order $(\eps_{\bar 5_1}/\eps_{\bar
5_2})^2$ \cite{Babu:tenc}, much larger than in typical GUTs.

In models where the CFT breaks baryon number, we expect some
dimension-six baryon-violating operators in the K\"ahler potential to
be enhanced by the CFT.  This is simply because they are not inhibited
by any symmetry. These operators are problematic because they are
suppressed by a factor of $16 \pi^2/M_c^2$ and by nothing else, unlike grand
unification where they are suppressed by $4 \pi
\alpha_{GUT}/M_{GUT}^2$.  This makes it clear that if baryon number is
violated by strong dynamics, then $M_c$ cannot be far below $M_{GUT}$.
If such operators do cause observable baryon number violation, the
flavor structure of the branching ratios can give interesting insight
into the underlying flavor physics \cite{Babu:tenc}.  In particular,
our scenario will appear quite different from standard grand
unification, since only a subset of the possible operators will be
consistent with the approximate R-symmetry of the CFT.

\section{Explicit Models of Flavor}

In this section we examine several explicit examples of flavor
models. 
We find that   it is not difficult to find fairly simple
supersymmetric examples of our mechanism. The main model building issues which arise
are:
\begin{enumerate}
\item{\bf A Graceful Exit:} 
Our mechanism requires the existence of a new
gauge group and matter transforming under both new and standard gauge 
symmetries.  It is therefore necessary for the CFT to be perturbed by
small relevant operators which give mass to all colored or
electrically charged states and decouple
the CFT from all standard model fields at low energies.
It is also important that the theory should not be driven into a phase
where any standard model gauge symmetries are broken. In most cases
the spectrum will include vector-like composite states carrying
standard model  quantum numbers. In some cases these states are parametrically
lighter than $M_c$. Mixing between such states and the
quarks and leptons may significantly reduce order of magnitude
predictions for some of the Yukawa couplings. This may actually
improve the agreement of the models with reality, but complicates the
discussion.
\item{\bf Proton Decay:}
In some cases the couplings of the CFT may
violate baryon number, and the graceful exit will be required to occur
above $\sim 10^{15}$~ GeV in order to avoid proton decay from
dimension-six operators in the effective Kahler potential. Obtaining a sufficiently large flavor
hierarchy  by then requires that the theory reaches the vicinity of a 
fixed point at a scale of order $\sim 10^{17}$ GeV or above,  
and generates an anomalous dimension near 1
for some first generation particles. Avoiding proton decay from
dimension-five operators places the additional constraints discussed in
section~\ref{decay}.
\item{\bf Landau poles:} Our flavor models require a large number of
new fields carrying standard gauge charges. This need not be in
conflict with conventional coupling constant unification, as explained
in section \ref{unification}.  However it is necessary to avoid
increasing the beta functions for standard gauge couplings too much,
or they will reach a Landau pole below the quantum gravity scale. This
constraint is fairly weak if most of the new matter is not much
lighter than the scale of grand unification, since the gauge
contribution to the beta function is large and negative above this
scale.  However, one must take into account that the one-loop coefficient
of the standard model beta functions is nonperturbatively corrected,
as we discuss in section \ref{unification}.
The constraint of Landau poles may be evaded if the
standard model sector has a sensible Seiberg-dual description above
the putative pole.

\item{\bf Acceptable Superpartner Masses:} The coupling to the
conformal sector has a strong effect on the masses of the scalar
superpartners of the light quarks and leptons, and affects solutions
of the supersymmetric flavor problem.  A simple possibility is that
superpartner masses are generated through low-energy gauge mediation
\cite{Giudice:1998bp}. In this case it is required that the graceful
exit occurs above the messenger scale (which could be as low as 10 TeV
\cite{Nelson:1998gp,Cheng:1998nb}); otherwise our mechanism will undo the
flavor degeneracy which is characteristic of gauge mediation.
Generic supergravity models have serious flavor problems; however, in
specific realizations of our flavor mechanism, these may actually be
remedied by the strong dynamics.  We will discuss the form of this
remedy and the associated constraints on model building in section
\ref{flavorprob}, and more completely in a future paper \cite{future}.
\end{enumerate}

It is straightforward to build models which appear to have all of these
features.  However, the simplest models typically have ambiguities,
such as an unknown R-charge, which makes their dynamics uncertain and
limits their predictions.  In some of the following models we have
sacrificed elegance in favor of control and predictivity.  More
general models, in which the rough predictions of section
\ref{generic} survive, but which put only partial constraints on the
$\eps$ suppression factors, are both common and physically reasonable.

\subsection{A Simple Model for the Intergenerational Hierarchy based on
an $SU(3)^3$ GUT}
\label{suthree}
In this subsection we give a simple example of a supersymmetric model which explains the
intergenerational mass hierarchy. 
The Standard Model is embedded in the grand unified gauge group
$SU(3)^3\otimes \ZZ_3$ \cite{Rizov:1981dp,Glashow:1984gc}. This is a
subgroup of the more famous $E_6$ \cite{Gursey:1976ki},
with identical successful predictions for coupling constant unification. The
first $SU(3)$ becomes the color group,  the electroweak
$SU(2)$ is embedded in the second $SU(3)$, and the hypercharge $U(1)$
is contained in both  the second and third
 $SU(3)$'s.  
 The quarks and leptons are contained in three copies of the
27-dimensional representation
\beq
27\equiv(3,\bar 3,1)+(\bar 3,1,3)+(1,3,\bar 3)\ .
\eeq
This representation also contains exotic fields; however, 
these are vector-like
under the standard model gauge group and may obtain mass at the GUT scale.

The standard model Higgs fields, and other Higgs fields needed to break
$SU(3)^3$ down to the standard model, may be placed in a couple of
additional $27_H+\overline{27}_H$ representations.
For arbitrary  superpotential couplings, this structure results in
unconstrained quark and lepton Yukawa coupling matrices to the light Higgses.

Possible Lagrangians for the Higgs sector are weakly coupled
and have already been discussed in the literature 
\cite{He:1986cs,Babu:1986gi,Nishimura:1988fp,Carlson:1992ew,Wang:1992hu,Lazarides:1993sn,Dvali:1994wj,Lazarides:1995px} so here we
confine our discussion to the strongly-coupled flavor sector.

The full gauge symmetry of the model is assumed to be 
$SU(3)^3\otimes \ZZ_3\otimes SU(4)\otimes SU(5)$, 
with matter in the representation given in Table~1.
\vskip .2 in 
\begin{tabular}{|l|c|c|c|c|}
\hline
\hfil  &\hfil $SU(3)^3$ \hfil &\hfil 
$SU(4)$\hfil &\hfil $SU(5)$ \hfil&dimension \\ \hline
&&&&\\ [-8pt]
$27_1,\;27_2,\;27_3$&$(3,\bar3,1)+(1,3,\bar3)+(\bar3,1,3)$&1&1&${5\over3},{4\over3},1$\\
$27_H, 27_H'$&$(3,\bar3,1)+(1,3,\bar3)+(\bar3,1,3)$&1&1&$1,1$\\
 $\overline{27}_H, \overline{27}_H'$&$(\bar3,3,1)+(1,\bar3,3)+(3,1,\bar3)$&1&1 &1,1 \\ 
 $Q$  &$(3,1,1)+(1,3,1)+(1,1,3)$&4&1&${5\over6} $            \\ 
 $\bar Q$&$(\bar3,1,1)+(1,\bar3,1)+(1,1,\bar3)$&$\bar4$&1 &${5\over6}$  \\ 
$Q'$  &$(3,1,1)+(1,3,1)+(1,1,3)$&1&$5 $ &${2\over3}$           \\ 
 $\bar Q'$&$(\bar3,1,1)+(1,\bar3,1)+(1,1,\bar3)$&1&$\bar5$&${2\over3}$ \\ 
\hline
\end{tabular}
\vskip .2 in 
Table 1.
{\it Quantum numbers and scaling dimensions at fixed point of chiral superfields in the model.   }
\vskip .2 in
        It has been shown \cite{Seiberg:1995pq} that   $SU(4)$ and $SU(5)$  
supersymmetric gauge theories with 9 flavors will flow to a 
superconformal phase in the infrared. The $SU(3)^3$ gauge groups  
are not asymptotically free, and their couplings will remain weak and have no effect on the
dynamics. The $SU(3)^3$ GUT symmetry breaking is done via a conventional weakly
coupled superpotential involving the various $27$'s. In addition, we assume
gauge-invariant and $\ZZ_3$-invariant superpotential couplings
\beq
{{\sum}}_{i=1,2}\lambda_i \bar Q Q 27_i +\bar \lambda'\bar Q' Q'
27_1\ .
\eeq
Note that this is the most general gauge-invariant trilinear
superpotential coupling
allowed between the $Q,Q',\bar Q,\bar Q'$ fields and the quarks and
leptons of the three generations. Without loss of generality, the field
$27_3$ then
has no relevant couplings to the superconformal sector.  Since its
Yukawa couplings will therefore be unsuppressed, it contains the
fermions which we will identify, after the fact, with the third
generation.
 
 We describe the assumed
dynamics of the theory
 from the top down in energy scale. First,  the
$SU(5)$ gauge coupling becomes strong at a scale $\Lambda_5$ and the theory 
flows to the
vicinity of the infrared fixed point found by Seiberg
\cite{Seiberg:1995pq}. At this fixed point, the $\bar Q',Q'$ fields obtain
anomalous dimensions of $-1/3$, and the $\lambda'$ coupling becomes
relevant. This coupling quickly becomes strong and we assume
it drives the theory to
the  vicinity of a new fixed point. Here
the field $27_1$ must acquire  a positive anomalous dimension of
$2/3$, and the
coupling
$\lambda_1$ as well as the couplings of the $27_1$ field to the Higgs fields
become irrelevant and highly suppressed. 

The $SU(4)$ gauge coupling becomes strong at a different scale
$\Lambda_4$ (which could be larger than $\Lambda_5$, but let us take it to
lie near or below $\Lambda_5$ for the present discussion.)  Below this
scale the $SU(4)$ gauge dynamics are also superconformal, with the
$\bar Q, Q$ fields acquiring anomalous dimensions of $-1/6$. The
coupling $\lambda_1$ remains weak and irrelevant, but the coupling
$\lambda_2$ is relevant and will drive the theory to a fixed point
where the anomalous dimension of the field $27_2$ is $+1/3$.

For several energy decades the theory remains at this fixed point
with the couplings $\lambda'$ and $\lambda_2$ strong and scale invariant and
all other superpotential couplings weak. A perturbative
 superpotential for the $27_H$ and $\overline{27}_H$ fields will
break the $SU(3)^3$ to the standard model at  a scale
$M_G\sim 10^{16}$~GeV. However the $SU(3)^3$ remains an approximate
global symmetry of all strong couplings of the theory. 
The scales $\Lambda_{4,5}$ could be higher or
lower than $M_G$ without significantly affecting the analysis.

Eventually, relevant interactions drive the theory out of the
superconformal phase, leaving an effective theory which is the
MSSM. These interactions could most simply be mass terms $ m_4 \bar Q
Q+ m_5 \bar Q' Q'\ .$ However if $m_{4,5}$ are fundamental parameters
it is hard to understand why they should be much smaller than the
fundamental scale $M$ of the theory.  A more attractive model is to
add another strong group $G$, with no matter fields and gauge field
strength $W$, and nonrenormalizable couplings 
\beq \int d^2\theta
\CO{1\over M^2} W^2(\bar Q Q+ \bar Q' Q') .
\eeq
Condensation of the $G$ gauginos would then generate small effective
mass terms for the $Q,\bar Q, Q',\bar Q'$ fields and drive the theory
to a phase where $SU(4)$ and $SU(5)$ are confining. 

 The standard-model-singlet operators $\bar QQ$ and $\bar Q' Q'$
develop expectation values which break $SU(4)$ and $SU(5)$.  Although
these may lie above the scale $\vev{W^2}/M^2$, they can be arranged to
lie well below the scale at which the CFT is reached, so the conformal
regime is not eliminated.  
 
The main effect of the superconformal dynamics on low energy
phenomenology is to generate large wave function renormalizations for
two generations of quarks and leptons. This will result in substantial
suppression of all superpotential couplings for the fields of the
first and second generations, with $\eps_1\sim \eps_2^2$ if
$\Lambda_4\sim\Lambda_5$ and $m_4\sim m_5$. In this case the
suppression factors are universal within a generation. Thus while the
intergenerational mass hierarchy and mixing pattern are roughly
explained, this theory makes the prediction that $m_u/m_t$ should be
similar to $m_d/m_b$, if all Yukawa couplings are initially
random. This prediction fails by about two orders of magnitude. The
suppression factors also do not explain the size of $m_b/m_t$.  Thus
in this model not all aspects of the flavor hierarchy are explained by
the renormalization group alone.  Obtaining a completely realistic
fermion mass spectrum is possible, however, since both mixing of
standard model fields with the composites $\bar Q Q$ and $\bar Q' Q'$
and mixing with the exotic down-type quarks and leptons will affect
the low-energy Yukawa couplings.  The ratio $m_b/m_t$ could either be
due to large $\tan\beta$ or to the location of the light down-type
Higgs within the $27_H+\overline{27}_H$ Higgs multiplets; for
instance, if the down-type Higgs were mostly a mixture of the
$\overline{27}_H$ fields, its couplings to the quarks and leptons
would be suppressed, since only a $27_H$ can couple to quarks and
leptons. Detailed model building in this direction will not be
attempted here.

\subsection{A ``10-centered'' model without proton decay}
\label{tencentered}
Here we give an example of a ``10-centered" model, {\it i.e.} a model  
which is compatible with $SU(5)$ grand unification and which produces suppression factors
for  two $SU(5)$ decuplets. Note that the conformal sector couplings do not
induce proton decay and so $M_c$ may be as low as $\sim $ 10
TeV.  Explaining $m_b/m_t$ will require either adding another
conformal sector
to suppress the couplings of the $\bar 5$'s, large $\tan\beta$, or a
mechanism to suppress the couplings of $H_d$. The gauge group and
field content of the model are listed in table 2.  

\vskip .2 in 
\begin{tabular}{|l|c|c|c|c|}
\hline
\hfil  &\hfil $SU(5)_{\rm GUT}$ \hfil &\hfil $Sp(8)$\hfil &\hfil $Sp(8)'$ \hfil&dimension \\ \hline
&&&&\\ [-8pt]
$T_{1,2,3}$&$10$&1&1&${42\over 25},{69\over 50},1$\\
$\bar F_{1,2,3},\bar H$&$\bar 5$&1&1&1\\
 $H$&$5$&1&1&1  \\ 
 $Q$  &$\overline{10}$&8&1&${87\over100}$             \\ 
$A$  &1&27&$1 $ &${3\over5}$           \\ 
 $J,K,L,M$&$1$&$8$&$1$&${3\over4},{3\over4},{3\over4},{9\over20}$\\ 
 $\bar Q'$&$10$&$1$&8 &(confined)  \\ 
 $R,S$    &1&1&8&(confined)\\
\hline
\end{tabular}
\vskip .2 in 
Table 2.
{\it Quantum numbers and scaling dimension of chiral superfields in the 10-centered model.   }
\vskip .2 in
We assume that   a $\ZZ_2$ symmetry exchanges $J$ and $K$
and that the first $Sp(8)$ group flows to a fixed point where the
following superpotential terms are marginal: 
\beq
W=T_1 Q L + T_2 Q M + A^5+ (JK)(JK)+ A^3L M+ (MJ) (MK) \ .\eeq

Relevant terms such as $A^3$ and $A^4$ must either be excluded by a
discrete symmetry or be initially extremely small, so as not
to disturb the approximate fixed point until at least the scale $M_c$.  
Since these would be forbidden by the
exact R-symmetry  which appears in the limit where the standard model
interactions are turned off, this is technically natural.
Small mass terms for  $L$, $M$ and $A$ may be added which will drive
the first $Sp(8)$ away from the fixed point into a confining phase. 
The $Sp(8)'$ is in a confining phase as well. The composite particles
are in a vector-like representation of the standard model 
gauge group and perturbatively
irrelevant couplings between the two sectors will allow all exotic
 particles to get masses.

The model gives $T_1$ and $T_2$ large anomalous dimensions of $17/25$
and $19/50$ respectively. The resulting prediction for the suppression
factors
\beq \eps_{10_1}=\eps_{10_2}^{34/19}
\eeq
is in good agreement with  \eref{epsilons}.

\subsection{Another $SU(5)$ based example}
\label{another}
The following model is also consistent with $SU(5)$ grand unification,
and produces suppression factors for two $10$'s and a $\bar 5$ from a
single $Sp(12)$ gauge group at a superconformal fixed point. A $\ZZ_2$
symmetry (either gauged or global) is also assumed.  Either large
$\tan\beta$ or another sector is required to suppress the bottom quark
mass.  The marginal couplings of the theory do violate baryon number
and will lead to proton decay from dimension-six operators in the Kahler
potential. Thus an acceptable proton lifetime will require exiting the
conformal regime at a scale above $10^{15}$ GeV.  The field content of
the model is given in table 3.
\vskip .2 in 
\begin{tabular}{|l|c|c|c|c|}
\hline
\hfil  &\hfil $SU(5)_{\rm GUT}$ \hfil &\hfil $Sp(12)$\hfil &\hfil $\ZZ_2$ \hfil&dimension \\ \hline
&&&&\\ [-8pt]
$T_{1,2,3}$&$10$&1&1&$2,\frac43,1$\\
$\bar F_{1,2,3},\bar H$ & $\bar 5$ &1 &1& $\frac53,1,1,1$\\
 $H$&$5$&1&1&1  \\ 
 $\bar{T}$  &$\overline{10}$&12&1&$\frac23$             \\ 
$A$  &1&65&$1 $ &$\frac23$           \\ 
 $F$&$5$&$12$&$1$&$1$\\ 
 $Z,U,V$& $1$&$12$&$1,-1,-1$ &$\frac13,\frac76,\frac76$  \\ 

\hline
\end{tabular}
\vskip .2 in 
Table 3.
{\it Quantum numbers and scaling dimension of chiral superfields   }
\vskip .2 in
The scaling dimensions listed in the table follow from the assumption
that the theory flows to a fixed point where the following
superpotential terms are marginal:
\beq
W=T_1 \bar T Z+ T_2 \bar T Z A + {\bar F_1} F Z+ {\bar T}^3 F+ \bar T F F Z + A U V +
Z^2 U V + Z^2 U^2 + Z^2 V^2\ .\eeq

Small relevant mass terms for the fields $A$, $U$ and $V$ can
eventually drive the $Sp(12)$ into a confining phase. In this phase
the exotic fields carrying standard model quantum numbers are in a
vector-like representation and will obtain mass from the
superpotential.  Note that the exit is graceful.  The coupling of
$T_2$ to the CFT is removed when $A$ becomes massive, while $(TZ)$ and
$(FZ)$ are both massive due to the $TFFZ$ term in the superpotential.
After some order-one mixing of $T_1$ with $FZ$ and $\bar F_1$
with $TF$, the predictions for $T_1$ and $F_1$ are qualitatively
unchanged.

\subsection{A less predictive ``10-centered'' model}
\label{less}
Here we give another example of a 10-centered model, comparable to that
of section 6.2.  It has the advantage of being more compact, and
the disadvantage of being less predictive, although potentially
just as realistic.

\vskip .2 in 
\begin{tabular}{|l|c|c|c|c|}
\hline
\hfil  &\hfil $SU(5)_{\rm GUT}$ \hfil &\hfil $Sp(8)$\hfil &\hfil $Sp(8)'$ \hfil&dimension \\ \hline
&&&&\\ [-8pt]
$T_{1,2,3}$&$10$&1&1&?,?,1\\
$\bar F_{1,2,3},\bar H$&$\bar 5$&1&1&1\\
 $H$&$5$&1&1&1  \\ 
 $Q$  &$\overline{10}$&8&1&?    \\   
 $L,M$&$1$&$8$&$1$&?,? \\   
 $J_1,J_2,J_3,J_4,J_5,J_6$&$1$&$8$&$1$&?,?,?,?,$\frac34,\frac34$\\ 
 $\bar Q'$&$10$&$1$&8 &(confined)  \\ 
 $\bar J'_1,J'_2$&$1$&$1$&8 &(confined)  \\ 
\hline
\end{tabular}
\vskip .2 in 
Table 4.
{\it Quantum numbers and scaling dimension of chiral superfields in the
model.   }
\vskip .2 in
We assume that  the first $Sp(8)$ group flows to a fixed point where the
following superpotential terms are marginal: 
\beq
W = (J_1J_2)^2+(J_3J_4)^2+(J_5J_6)^2+  
(LJ_1)(J_1J_3)+ T_2 Q M + T_1 Q L 
\eeq
This superpotential can be assured by a gauged $\ZZ_8$ discrete
symmetry under which $L$ and $M$ carry identical charges. The fourth
term in the superpotential defines $L$, without loss of generality;
the term $(MJ_1)(J_1J_3)$ may be removed by a rotation of $L$ and $M$.
The last two terms define $T_1$ and $T_2$, where we have guessed that
$M$ will have larger dimension than $L$.  Graceful exit may occur
through masses for $J_1,J_2,J_3,J_4$, $L$ and $M$, after which the
$Sp(8)$ gauge group confines and (through small couplings such as
$QQ\bar Q'\bar Q',\dots$) all exotic fields become massive.

Notice that this model provides insufficient constraints to determine
the R-charge of the CFT, and therefore we do not know the dimensions
of most operators.  However, the symmetry between $L$ and $M$ is broken, so
$T_1$ and $T_2$ will have different anomalous dimensions.

This model may be altered by using other $Sp$
gauge groups and changing the number of fields in the fundamental
representation.  It is likely that at least one of these models gives
suppression factors which are consistent with data.

\subsection{Suppression factors for the ${\bf \bar 5}$ fields.}

Ideally a simple model would suppress all standard model
fields at once.  However, it is straightforward to suppress those
of the ${\bf \bar 5}$ fields separately.  Let us consider
a particularly simple (although not fully predictive) model.
This model also is useful for illustrating another means by which a hierarchy
in the suppression factors may be obtained.

\vskip .2 in 
\begin{tabular}{|l|c|c|c|}
\hline
\hfil  &\hfil $SU(5)_{\rm GUT}$ \hfil &\hfil $Sp(4)$ \hfil&dimension \\ \hline
&&&\\ [-8pt]
$T_{1,2,3}, T_0$&$10$&1&1\\
$\bar F_{1,2,3,4},\bar H$&$\bar 5$&1&?,1\\
 $H$&$5$&1&1  \\ 
 $X$  &$5$&4&?    \\   
 $R_{1,2,3,4,5}$&$1$&$4$&?\\ 
\hline
\end{tabular}
\vskip .2 in 
Table 5.
{\it Quantum numbers and scaling dimension of chiral superfields that
lead to suppressions of $F_i$ fields.   }
\vskip .2 in

The superpotential is 
\beq
W= \bar F_i XR_i + T_0 XX 
\eeq
One linear combination (let us call it $R_5$) of the $R_i$ 
does not couple to the $\bar F_i$.  It is easy to prove that
$ \dim (XR_i) < 2$; that $\dim (X R_i) > 1$ is not proven but
is extremely plausible, since the one-loop coefficient
of the $Sp(2)$ gauge-coupling beta function is not
large.  By symmetry the four $\bar F_i$ have
equal and positive anomalous dimensions. 

Let us assume that at some scale a dynamical mass matrix $m^{ij}$ is
generated for the five fields $R_i$.  This removes
four of the $R_i$ and leaves one $\bar F_i$ coupled to one linear
combination, call it $R_0$, of the $R_i$.  Confinement of $Sp(2)$ now
occurs; $XR_0$ becomes massive with a linear combination of the 
$\bar F_i$, leaving three $\bar F_i$ behind to make up standard model matter.
$T_0$ and $XX$ are also massive.

To obtain a small hierarchy in the $\epsilon_{\bar 5_i}$, as in
\Eref{epsilons}, is not difficult.   Suppose that the couplings
$\lambda_i$ in front of the $F_i X R_i$ operators are somewhat smaller
than their CFT values and are slightly hierarchical
at the Planck scale.  They are relevant and will grow to be
strong, but since $\beta(\lambda_i) \propto \lambda_i$, they
will do so in a hierarchical manner, so that one, let us call it
$\lambda_1$, might become strong before the others.  In this
case $\epsilon_{\bar 5_1}$ will be slightly smaller than
the other suppression factors, as preferred phenomenologically.

\section{The Supersymmetric Flavor Problem}
\label{flavorprob}
A second attractive, although not strictly necessary, feature of this
scenario is that it can solve the supersymmetric flavor problem.  This
will be spelled out in detail in a future paper 
so we will only summarize the mechanism here.  No such solution is
necessary for gauge-mediated and some anomaly-mediated models, so let
us assume we have a typical supergravity model of supersymmetry
breaking, in which the contributions of the soft supersymmetry
breaking terms to flavor-changing neutral currents (FCNCs) and
electric dipole moments (EDMs) are generally unsuppressed. Exact
relations for the renormalization of soft supersymmetry breaking terms
\cite{future,Hisano:1997ua,Avdeev:1997vx,Jack:1997pa,Jack:1998eh,Jack:1998iy,Karch:1998xt,Karch:1998qa,Kobayashi:1998jq}
can be used to show that in our flavor scenario, the same anomalous
dimensions which reduce the fermion Yukawa couplings suppress the
trilinear scalar couplings (A-terms) between squarks/sleptons and the
Higgs bosons.  The A-terms therefore have a hierarchy similar to the
Yukawa couplings, which reduces FCNCs and EDMs close to experimental
bounds \cite{Masiero:1999ub}, provided that $\tan\beta$ is not large.

Meanwhile, as shown in
\cite{Karch:1998qa,Luty:1999qc} and established generally in
\cite{future}, under certain special circumstances a CFT to which
neutral fields $X_i$ are coupled can drive the soft masses of the
$X_i$ to zero. This happens whenever the anomalous dimension of the
standard model fields can be uniquely determined from an  R-symmetry
contained within the superconformal algebra.  For instance, this 
effect occurs, for all fields obtaining suppression factors,
in the  examples discussed in sections
\ref{suthree}---\ref{another}. Driving the soft
masses of the superpartners of the light quarks and leptons to zero
would be a phenomenological disaster if it occurred at the weak scale,
but if it occurs at a high scale $M_c$, then the renormalization
between the scales $M_c$ and $M_W$ will affect these masses
significantly.  In particular, since the gauginos are not coupled to
the CFT, gaugino masses are not driven small.  The gaugino masses then
give positive and flavor-blind additive contributions to the light
squark and slepton masses, as has been used to advantage in ``no-scale''
supergravity models \cite{Cremmer:1983bf,Ellis:1984bm}.  
These contributions are logarithmically 
enhanced, and any
flavor-violating effects, suppressed to near zero by the CFT, are
dwarfed by the time the TeV scale is reached. 


\EPSFIGURE{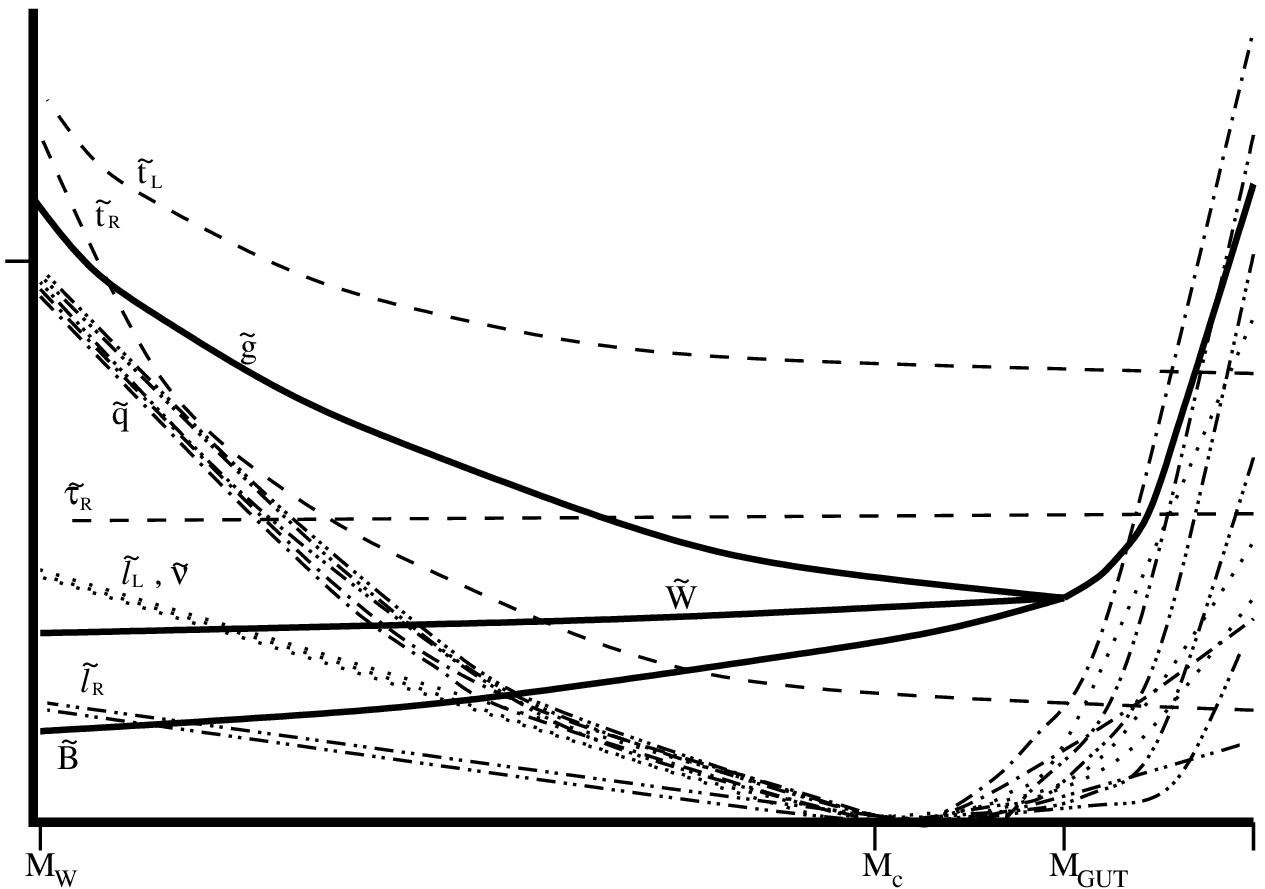}{Schematic possible renormalization group flow
for the gaugino and sfermion masses.  The gauginos $\tilde g,\tilde
W,\tilde B$ have masses which scale with the standard model gauge
couplings.  The sleptons $\tilde \ell_L,\tilde \nu$ end up degenerate,
as do the $\tilde \ell_R$ sleptons and the squarks $\tilde q$, with
the exception of some sfermions of the third generation. }


Thus, the squarks and sleptons of the first two generations (and
possibly the right-handed bottom squark and one or both tau sleptons)
end up with degenerate masses; the remaining squarks and sleptons have
masses which are sensitive to the Planck scale and are not predicted
by the scenario. That suppression of
scalar masses at a high scale ameliorates the supersymmetric flavor problem has been
noted previously \cite{Choudhury:1995pn}; however our mechanism has the
distinctive feature that not all scalar masses are 
suppressed. 
The qualitative behavior of the masses as a function
of scale is illustrated in the figure.  Quantitatively, we find
that the masses at low scales are determined by a D-term for $U(1)$
hypercharge (which can be naturally generated from the soft $H_u$ and
$H_d$ supersymmetry breaking masses) as well as by $M_c$ and by the
gaugino masses.  Working to one-loop in the standard model
couplings, we find the first and second generation squarks have
masses $\tilde m_{q}^2$ which are related to the gluino mass
$M_{\tilde g}$ by
\def\tim{\tilde m^2}
\beq
{\tim_{q}\over M_{\tilde g}^2}\approx {8\over 9}\left[1 -
{\alpha_3(M_c)^2\over \alpha_3^2}\right] \ .
\eeq
(Here and in the following, $\alpha_i$ with no argument is evaluated
at or slightly above $M_W$.) Since this is not very sensitive to
$M_c$, it is a prediction of the model that $\tilde m_{q}/|M_{\tilde
g}| \ltap .9$.  The $U(1)$ D-term makes predictions for the sleptons
more uncertain, but cancels out in the linear combination $\tim_\nu +
\tim_{e^-} + \tim_{e^+}. $ If we assume the gauginos have the same
mass at or near the GUT scale, we find the average mass-squared for the
sleptons is
\beq
{\tim_\nu + \tim_{e^-} + \tim_{e^+} \over 3M_{\tilde g}^2 } \approx 
\left[{\alpha_2^2\over \alpha_3^2}\right]
\left[{\alpha_2^2(M_c)\over \alpha_2^2}-1\right]
+
{5\over 66}
\left[{\alpha_1^2\over \alpha_3^2}\right]
\left[{\alpha_1^2(M_c)\over \alpha_1^2}-1\right] \ .
\eeq
This ratio is about $(1/4)^2$ if $M_c\sim M_{GUT}$, decreasing to
$(1/6)^2$ if $M_c\sim 10^{11}$ GeV.  Note that each of these two
observables gives a (rather imprecise) measurement of $M_c$.
Furthermore, at one loop the quantity $2\tim_q - \frac13\tim_{\bar u}
- \frac53\tim_{\bar d}$ is independent both of the gluino contribution
and of the hypercharge D-term contribution.  The hypercharge fermion
loop is small, so the wino graph dominates for this
observable. Numerically the left-handed squarks will be heavier than
the right-handed squarks by order 30-40 GeV, if $M_{\tilde g}\sim 1$
TeV.  This gives yet another test of the model, and an independent
rough measure of $M_c$.

Since the hypercharge gaugino has a mass of order $ M_{\tilde g}/6$, it cannot
be the lightest supersymmetric particle (LSP) unless $M_c$ is around
$10^{11}$ GeV or above.  Even for $M_c$ of order $M_{GUT}$, the
hypercharge D-term must be positive, raising the masses of the
electroweak-singlet sleptons relative to the doublet sleptons, if the
the hypercharge fermion is to be the LSP.  For this purpose, the
D-term must be reasonably large, of order $(100$ GeV$)^2$. 

If the slepton masses are of order 200 GeV and those of the squarks
are of order 800 GeV, then typically the only phenomenological
difficulties come from lepton-sector A-terms in the electric dipole
moment of the electron and in $\mu\to e\gamma.$  These are not far from
the experimental bounds, however, so a small additional systematic or
accidental suppression of these A terms will be enough to make the
scenario viable. 

\section{Coupling Constant Unification and Fermion Mass Relations}
\label{unification}

Any flavor model involving new strong dynamics can  have an
important effect on coupling constant running, potentially spoiling
the coupling constant unification of the MSSM. However, as was
explained in \cite{Nelson:1998gp}, coupling unification is
retained if the strong dynamics respects a global symmetry
into which the standard model gauge symmetry
can be embedded, as long as that global group would force unification
if it were gauged.  Examples include $SU(5)$ or $\ZZ_3\times SU(3)^3$
global symmetries.
  
The proof is quite simple, and holds to one loop in the standard model
couplings and to all orders in other couplings.  According to
\cite{Shifman:1986zi,Shifman:1991dz} the coupling $g_k$ runs as
\be\label{SVbeta}
\beta_{g_k} = -{g_k^3\over 16 \pi^2}
{3C_2(G_k)-\sum_p T_k(\phi_p)[1-\gamma_{\phi_p}]\over
1-{g_k^2\over 8\pi^2}C_2(G_k)} \ .
\ee
Here $C_2(G_k)$ is the second Casimir operator of the gauge group
$G_k$ for which $g_k$ is the coupling, the sum in the numerator is
over all matter fields $\phi_p$, $T_k(\phi_p)$ is half the index of
the representation of $\phi_p$ under $G_k$, and $\gamma_{\phi_p}$ is
the anomalous dimension of $\phi_p$.  To leading order in a weak
coupling constant $g_k$ the beta function is proportional simply to $g_k^3$
times $3C_2(G_k)-\sum_p T(\phi_p)[1-\gamma_{\phi_p}]$.  The usual
statement of coupling constant unification is that a complete $SU(5)$
multiplet $\{\phi_j\}$ preserves unification because $\sum_j
T(\phi_j)$ is the same for each standard model group factor, leading
to equal shifts in $b_0 = 3C_2(G_k)-\sum_p T(\phi_p)$ for the three
groups and preserving both unification and the unification scale.  In
our case, the $SU(5)$ multiplets have large anomalous dimensions due
to strong interactions involving the CFT sector. However, since the
fields $\{\phi_j\}$ in each multiplet all have the same anomalous dimension
$\gamma_{\{\phi_j\}}$ (by approximate $SU(5)$ flavor symmetry,) the sum
$\sum_j T(\phi_j)[1-\gamma_{\phi_j}]= [1-\gamma_{\{\phi_j\}}]\sum_j
T(\phi_j)$ is essentially the same in each standard model group
factor.  Again, unification is preserved.

Note that {\it any} global symmetry which requires that
\be\label{condition}
\sum_p T_1(\phi_p)\gamma_{\phi_p}=\sum_p T_2(\phi_p)\gamma_{\phi_p}=
\sum_p T_3(\phi_p)\gamma_{\phi_p}\ 
\ee
will make standard predictions for the unification scale and coupling
constants. In addition, if \eref{condition} is satisfied  for the
standard model fields alone then one can make a prediction for fermion
mass relations.

The anomalous dimensions of  standard model fields are
related
to their Higgs Yukawa couplings through  
\be\label{ybeta}
\beta_{y_{ijk}} = {1\over 2} y_{ijk}
\left[\gamma_i+\gamma_j+\gamma_k\right]\ ,
\ee
which has solution
\be\label{yuksol}
{y_{ijk}(\mu)\over y_{ijk}(\mu_0)}=e^{-\int (\gamma_i+\gamma_j+\gamma_k)
dt}\ee
where $t=\ln(\mu_0/\mu)$.
Applying the constraint \eref{condition} to the anomalous dimensions of
the quarks and leptons gives the relation
\be
\sum_i (2\gamma_{q_i}+\gamma_{\bar u_i}+\gamma_{\bar e_i})=
\sum_i (3\gamma_{q_i}+\gamma_{ \ell_i})=
\frac65\sum_i\left(\frac16\gamma_{q_i}+\frac12\gamma_{ \ell_i}
+\frac43\gamma_{\bar u_i}+\frac13\gamma_{\bar d_i}+\gamma_{\bar e_i}\right)
\ ,
\ee
which implies that
\be\label{downlepton}
\sum_i(\gamma_{q_i}+\gamma_{\bar d_i})=
\sum_i(\gamma_{\ell_i}+\gamma_{\bar e_i})
\ .
\ee
 If we assume that all fermion Yukawa couplings start out at the same order
of magnitude, and neglect any anomalous
dimensions for the Higgs, then eqs. \eref{yuksol} and \eref{downlepton}
imply that, up to corrections from weak gauge couplings,
\be\label{massrel}
\ln{m_d m_s m_b}=\ln{m_e m_\mu m_\tau}\ .
\ee
When standard model gauge corrections (principally QCD) are included
then \eref{massrel} is  well satisfied.

\section{Non-supersymmetric CFT's and Flavor} 
\label{technicolor}

It may be possible to realize this mechanism for generating the
flavor hierarchy in a nonsupersymmetric theory. Of course, such a
theory must account for electroweak symmetry breaking. Since scalars
in such a theory are unnatural, it is more attractive to break the
electroweak symmetry dynamically, {\it e.g.}  through a Higgs
boson-like condensate which is a composite of two fermions
\cite{Weinberg:1976gm,Weinberg:1979bn,Susskind:1978ms}.  In this case
the Yukawa couplings of the fermions are in fact four-fermion
couplings above the weak scale.

Here we give a brief discussion of how a flavor hierarchy might be
generated, although it is not known whether the required dynamics can
actually occur.  Consider a theory consisting of the standard model
fields and a new sector which flows to a conformal fixed point. This
new sector may contain new fermion fields with four-fermion
interactions between them.  Note that four-fermion interactions,
irrelevant if the gauge couplings are small, can be marginal or
relevant when they are large, a point often used in technicolor models
\cite{Holdom:1981rm}.  The ordinary quarks and leptons are assumed not
to carry any non-standard gauge interactions. However, quarks and
leptons may couple to three-fermion operators of this CFT.  Unitarity
requires the quarks and leptons to have positive anomalous dimensions,
so for such couplings to be marginal or relevant the three-fermion
operators must have dimension less than $5/2$. On the other hand,
unitarity only constrains them to have dimension greater than $3/2$,
so there is no obvious obstruction to this possibility.  The quarks
and leptons which couple to these operators could therefore obtain
large positive anomalous dimensions.\footnote{This should be stated
more precisely.  First consider the theory in which the couplings
between the quarks and leptons and the three-fermion operators are
absent.  Assume that the theory flows to a CFT.  Now add the couplings
back as small perturbations.  If the dimension of the three-fermion
operators is greater (less) than $5/2$, then conformal perturbation
theory assures that these perturbations will be irrelevant (relevant).
If the couplings are relevant the theory may flow to a new CFT.  The
properties of the new CFT are non-perturbative in the four-fermion
couplings and cannot be inferred from the initial CFT.  Thus, although
there is no evidence that it is impossible, we cannot prove that the
above conditions sometimes lead to a CFT with large anomalous
dimensions for quarks and leptons.}

 As in the supersymmetric case studied above, the large anomalous
dimensions obtained in this way will suppress the four-fermion
interactions which lead to the quark and lepton Yukawa couplings, and
will thereby generate a flavor hierarchy.  However, unlike the
superconformal case, where certain scaling dimensions are proportional
to $U(1)_R$ charges and are therefore additive, nonsupersymmetric
theories do not have such a simple property.  The scaling
dimensions of products of light fermions are not the sum of the
scaling dimension of those fields.  Therefore, a simple relation for
Yukawa matrix scaling would be guaranteed only if a separate conformal
sector were introduced for each light field.

There is, however, a further dynamical concern.  If quarks and leptons
have only positive anomalous dimensions, then the fermion bilinear
which leads to the Higgs boson would need to have scaling dimension
close to unity, as in walking technicolor theories, in order to avoid
suppression of the top quark mass.  This could be avoided if the top
quark is part of the sector which generates electroweak symmetry
breaking, as in topcolor \cite{Hill:1991at}, or more generally, 
if the third generation
carries
any new strong gauge symmetry.  If the dimension of the Higgs
operator is sufficiently close to unity, then flavor physics can be
pushed up to high enough scales to decouple any flavor-changing
neutral current effects.  Such an anomalous dimension for the Higgs operator 
 might be natural in a large $N_c$
technicolor
theory
with matter content tuned so that the theory is nearly on the edge
between the conformal and chiral symmetry breaking phases --- assuming
these phases actually are adjacent \cite{Appelquist:1996dq}.

In
strongly-coupled nonsupersymmetric conformal theories there are few
known general constraints on anomalous dimensions, so we regard it to
be a completely open question whether a natural and viable theory of
fermion masses and electroweak symmetry breaking can be constructed without
fundamental scalars or supersymmetry.

\section{Comments on String Models}
\label{string}

Our anarchy-suppression scenario tells a cautionary tale, and perhaps
a suggestive one, for string model building.  The preconditions for
the scenario are natural in string theory.  Yukawa couplings may
easily be large, and all of the same order, at the
string scale.  Also, it is typical in string
models for additional non-abelian gauge groups to accompany the
standard model, and for there to be matter fields which are charged
under both the new groups and the standard model.  If all of these
groups start out with moderate gauge couplings, but the standard model
couplings are initially driven small due to positive beta functions,
the result is a weakly-coupled standard model coupled to a strongly
interacting sector.  The dynamics of this new sector may be very
complicated just below the compactification scale, but very often,
since conformal field theories are myriad in supersymmetric models,
its near-infrared behavior will be approximately conformal.  If (as is
typical in string models) some standard model fields couple linearly
via superpotential terms to fields charged under this new sector, they
will obtain large anomalous dimensions.  These will in turn have a
drastic effect on the standard model Yukawa couplings to the Higgs
boson, as we have described, as well as on the soft supersymmetry
breaking terms.

In such models, it is completely misleading to attempt to understand
flavor (and baryon-number violation, for that matter --- see
section \ref{decay}) using the
Lagrangian obtained just below the compactification scale.  Instead,
one must perform a detailed field-theoretic analysis, as we have done
here.  In a string compactification with the standard model (or a GUT)
along with an extra gauge group $G$ and extra matter $Q$, the
superpotential coupling the $Q$ fields to themselves and to the
standard model fields must be obtained.  Then, setting all standard
model gauge and Yukawa couplings to zero, one must ask whether the
gauge group $G$ can have an infrared fixed point.  A prerequisite for
such a fixed point is an R-symmetry, anomaly free under $G$, under
which all operators to have dimensions consistent with the unitarity
bound.  (One must approach this issue carefully, since the R-symmetry
may be an accidental symmetry visible only when certain terms in the
superpotential flow in the infrared to zero.)  If the R-symmetry is
unique, then the dimensions of the various operators in the putative
conformal field theory can be obtained and the rough effect on the
standard model Yukawa couplings estimated\footnote{Even if the
R-symmetry is not unique, there may still be special operators whose
dimensions are known, leading to a few testable predictions in the
masses and mixings of the standard model fermions.  For example, the
models of \cite{spinten} have one-parameter families of candidate
R-symmetries, but certain operators have R-charges which are
independent of the parameter.  Also, unitarity constraints lead to
inequalities which limit the range of the parameter.  If the standard
model is coupled to such a theory, the anomalous dimensions of quarks
and leptons may still be well enough known to rule out many string
models.}.  Only at this point can one determine whether the string
model is potentially viable.  Thereafter, one must ask the harder
dynamical question of whether the conformal sector can properly
decouple from the standard model without leaving over unacceptable
massless matter.

\section{Conclusions}
\label{conclusions}

We have demonstrated that coupling the standard model to a sector which
is nearly conformal for several decades in energy can explain the striking
features of the quark and lepton masses and mixing matrices. 
No flavor symmetry is required.  This is particularly
straightforward in supersymmetric theories, although the mechanism may
work more generally. In the supersymmetric case, a large mass
hierarchy and small mixing angles, with the largest mixing between
adjacent generations, are easily understood, as is a large mixing
angle for $\nu_\mu\leftrightarrow\nu_\tau$ oscillations. The
well-known successful predictions of supersymmetric grand unified
theories are not altered.  Meanwhile, $SU(5)$-based models (which may
or may not be GUTs) give successful order-of-magnitude predictions for
the fermion masses, as in previously studied ten-centered models
\cite{Babu:1996hr, Strassler:1996ia,Strassler:Nagoya,Babu:tenc}. 
Proton decay from dimension
five operators  can be naturally reduced below experimental bounds,
provided
$\tan\beta$ is not too large. 
Supersymmetric flavor-changing neutral currents are greatly suppressed
in a large class of models, perhaps to acceptable levels. This
mechanism of suppression of flavor-changing neutral currents makes
distinctive predictions for the superpartner spectrum, which will be
discussed in an upcoming paper \cite{future}.

\bigskip
\noindent\medskip\centerline{\bf Acknowledgments}
We gratefully
acknowledge the
Aspen Center for Physics, where part of this work was done.  M.J.S.
also thanks the Institute for Theoretical Physics at the
University of California Santa Barbara and the theory group
at the University of Washington for their hospitality. We thank Yossi
Nir for reading  the manuscript and making helpful comments.
A.E.N. is supported in part by DOE grant \#DE-FG03-96ER40956.
The work of M.J.S. was
supported by National Science Foundation grant NSF PHY95-13835 and by
the W.M. Keck Foundation.

\bibliography{flavor}
\bibliographystyle{JHEP}

\end{document}